\documentclass[
 reprint,
superscriptaddress,
,prl
]{revtex4-1}
\usepackage{xcolor}

\usepackage{graphicx}
\usepackage{dcolumn}
\usepackage{bm}
\usepackage{texshade}

\begin{document}

\preprint{APS/123-QED}

\title{Development of the Soft X-ray AGM-AGS RIXS Beamline at Taiwan Photon Source}

\author{A. Singh}
\author{H. Y. Huang}
\author{Y. Y. Chu}
\author{C. Y. Hua}
\author{S. W. Lin}
\author{H. S. Fung}
\author{H. W. Shiu}
 \author{J. Chang}
\affiliation{National Synchrotron Radiation Research Center, Hsinchu 30076, Taiwan}

\author{J. H. Li}
\affiliation{Department of Physics, National Tsing Hua University, Hsinchu 30013, Taiwan}

\author{J. Okamoto}
\author{C. C. Chiu}
\author{C. H. Chang}
\author{W. B. Wu}
\author{S. Y. Perng}        
\author{S. C. Chung}
\author{K. Y. Kao}
\author{S. C. Yeh}
\author{H. Y. Chao}
\author{J. H. Chen}
\affiliation{National Synchrotron Radiation Research Center, Hsinchu 30076, Taiwan}
     
\author{D. J. Huang}
\altaffiliation[Corresponding author:] {{
djhuang@nsrrc.org.tw}} 
\affiliation{National Synchrotron Radiation Research Center, Hsinchu 30076, Taiwan} \affiliation{Department of Physics, National Tsing Hua University, Hsinchu 30013, Taiwan}

\author{C. T. Chen}
\affiliation{National Synchrotron Radiation Research Center, Hsinchu 30076, Taiwan}

\date{\today}

\maketitle


\section{Abstract}
We report on the development of a high-resolution and highly efficient beamline for soft-X-ray resonant inelastic X-ray scattering (RIXS) located at Taiwan Photon Source.  This beamline adopts an optical design that uses an active grating monochromator  (AGM) and an active grating spectrometer (AGS) to implement the energy compensation principle of grating dispersion. Active gratings are utilized to diminish defocus, coma and higher-order aberrations as well as to decrease the slope errors caused by thermal deformation and optical polishing.  The AGS is mounted on a rotatable granite platform to enable momentum-resolved RIXS measurements with scattering angle over a wide range.  Several high-precision instruments developed in house for this beamline are briefly described.  The best energy resolution obtained from this AGM-AGS beamline was 12.4~meV at 530 eV, achieving a resolving power 42,000, while the bandwidth of the incident soft X-rays was kept at 0.5~eV.  To demonstrate the scientific impacts of high-resolution RIXS, we present an example of momentum-resolved RIXS measurements on a high-temperature superconducting cuprate, La$_{2-x}$Sr$_x$CuO$_4$. The measurements reveal the A$_{1g}$ apical oxygen phonons in superconducting cuprates, opening a new opportunity to investigate the coupling between these phonons and charge density waves.

\section{Introduction}

The energy dispersion of low-energy elementary excitations in momentum space reflects the fundamental physical properties of materials.  Resonant inelastic X-ray scattering (RIXS) is a powerful technique to probe these excitations with momentum resolution, and provides direct information about the dynamics arising from fluctuations of spin, charge and orbital degrees of freedom \cite{kotani2001, ament2011}.

The process of X-ray absorption in a material and its subsequent re-emission of an X-ray of different energy is known as inelastic X-ray scattering (IXS) \cite{schulke1989inelastic}. If the energy of the incident photons is tuned to an absorption resonance in which a core-level electron is excited to an unoccupied state, the subsequent X-ray emission spectrum depends strongly on the incident photon energy; this process is called resonant IXS, i.e., RIXS \cite{kotani2001, ament2011}.  It is also a scattering process in which the energy and momentum of the scattered X-ray conform to conservation rules, thus providing information about the energy and momentum of elementary excitations, such as $d$-$d$, charge-transfer, plasmon, magnon and phonon excitations etc. of quantum materials. The resonance effect significantly enhances the scattering cross section and offers a probe of elementary excitations with elemental and chemical selectivity. In addition, RIXS is a photon-in and photon-out technique that has been applied to explore matter in various phases.

Despite its unique advantages, RIXS was unpopular among available spectrometric techniques because of a lack of good energy resolution and a very weak signal intensity.  In the past decade, RIXS has, however, become widely accepted as one of the most powerful tools to investigate the properties of materials in terms of elementary excitations. After extensive developments of instrumentation, significant improvements in energy resolution and measurement efficiency have been achieved in the regime of soft X-ray energy.  For example, the energy resolution of the AXES monochromator and spectrometer \cite{dallera1996, ghiringhelli1998} improved from 500 meV at 530 eV in 1996 to 50 meV in 2013 \cite{ghiringhelli2012, schmitt2013, dinardo2007, ghiringhelli2006saxes} on replacing the microchannel plate detector with a charge-coupled device (CCD) detector  \cite{dinardo2007} and switching to a Dragon-type monochromator \cite{chen1987concept,chen1989}.  These improvements allowed researchers to study the $d$-$d$ excitations of 3$d$ transition-metal oxides with effective energy resolution \cite{ghiringhelli2004, ghiringhelli2006res}.  The SAXES spectrometer at Swiss Light Source \cite{ghiringhelli2006saxes} enabled measurements on magnetic excitations in cuprate superconductors \cite{braicovich2010, le2011intense}.  Furthermore, beamline ID32 at European Synchrotron Radiation Facility achieved resolution 30 meV at the Cu $L_3$-edge \cite{brookes2018beamline}.  This beamline enables RIXS intensity mapping as a function of momentum transfer through the rotation of the spectrometer in ultra-high vacuum (UHV).  A polarimeter has also been installed allowing the polarization analysis of scattered photons. Recent years have seen an enhanced development of new high-resolution soft X-ray RIXS instruments, including beamline I21 at Diamond Light Source, beamline SIX at National Synchrotron Light Source II \cite{jarrige2018paving}, the VERITAS soft X-ray RIXS beamline at MAX IV,  
the soft-X-ray spectrometer PEAXIS at BESSY II \cite{schulz2020} and  beamline 41A at Taiwan Photon Source (TPS) \cite{huang2018quest}.

To meet the two stringent requirements--high resolution and high efficiency--in soft X-ray RIXS experiments, a design concept of an active grating monochromator (AGM) and an active grating spectrometer (AGS) based on the principle of energy compensation of grating dispersion was conceived in 2002 and publised in 2004 \cite{fung2004}.  In the AGM-AGS design, the efficiency of RIXS measurements becomes greatly enhanced on increasing the bandwidth of the incident photons, while maintaining the energy resolution.  The energy-compensation principle for RIXS has been successfully tested at Taiwan Light Source (TLS) beamline 05A \cite{lai2014}.  Our theoretical simulations indicate that a resolving power better than $10^5$ for photon energies from 400~eV to 1000~eV is achievable with an AGM-AGS beamline, motivating us to build a new soft X-ray RIXS beamline at TPS \cite{huang2018quest}.

In this paper, we report on the development of the soft X-ray AGM-AGS RIXS beamline located at TPS port 41A.  This paper is organized as follows.  In Sec. 2, we introduce the RIXS branch of TPS 41A, including the design concept, a summary of precision instruments developed in house for this beamline and their performance.  In Sec. 3, we discuss commissioning results, including adjustment of the grating surface profile, resolution optimization and RIXS measurements on the phonon excitations of high-temperature superconducting cuprates (HTSC).  A summary and future plan follow in Sec. 4.

\section{RIXS of TPS beamline 41A }

\subsection{Photon source} 

Beamline 41A at TPS composes two branches, i.e., high-resolution RIXS and coherent soft X-ray scattering.  Both branches share the same monochromator, slits and front-end focusing optics.  The photon source originates from two elliptically polarized undulators (EPU) in tandem in a 12-m straight section with a double-minimum $\beta$ function to enhance the brilliance.  Each EPU magnet has length 3.2~m and period 48 mm.  The brilliance of the EPU tandem is designed to be greater than  $1\times10^{20}$ photons~s$^{-1}$mrad$^{-2}$mm$^{-2}$ per 0.1 $\%$ BW in the energy range from 400 eV to 1200 eV; the photon flux of the central cone exceeds $1\times10^{15}$ photons~s$^{-1}$.  In this energy range, the calculated beam sizes are about 386 $\mu$m and 28-35 $\mu$m at the full width at half maximum (FWHM) in the horizontal and vertical directions, respectively; the beam divergences are, respectively, $42-61~\mu$rad$ ~and~ $33-52$~\mu$rad at FWHM, in the horizontal and vertical directions, depending on photon energy. Detailed parameters of EPU 48 are reported elsewhere \cite{chang2010design, chung2016status}.

 \begin{figure}
\centering
\includegraphics[width=0.48\textwidth]{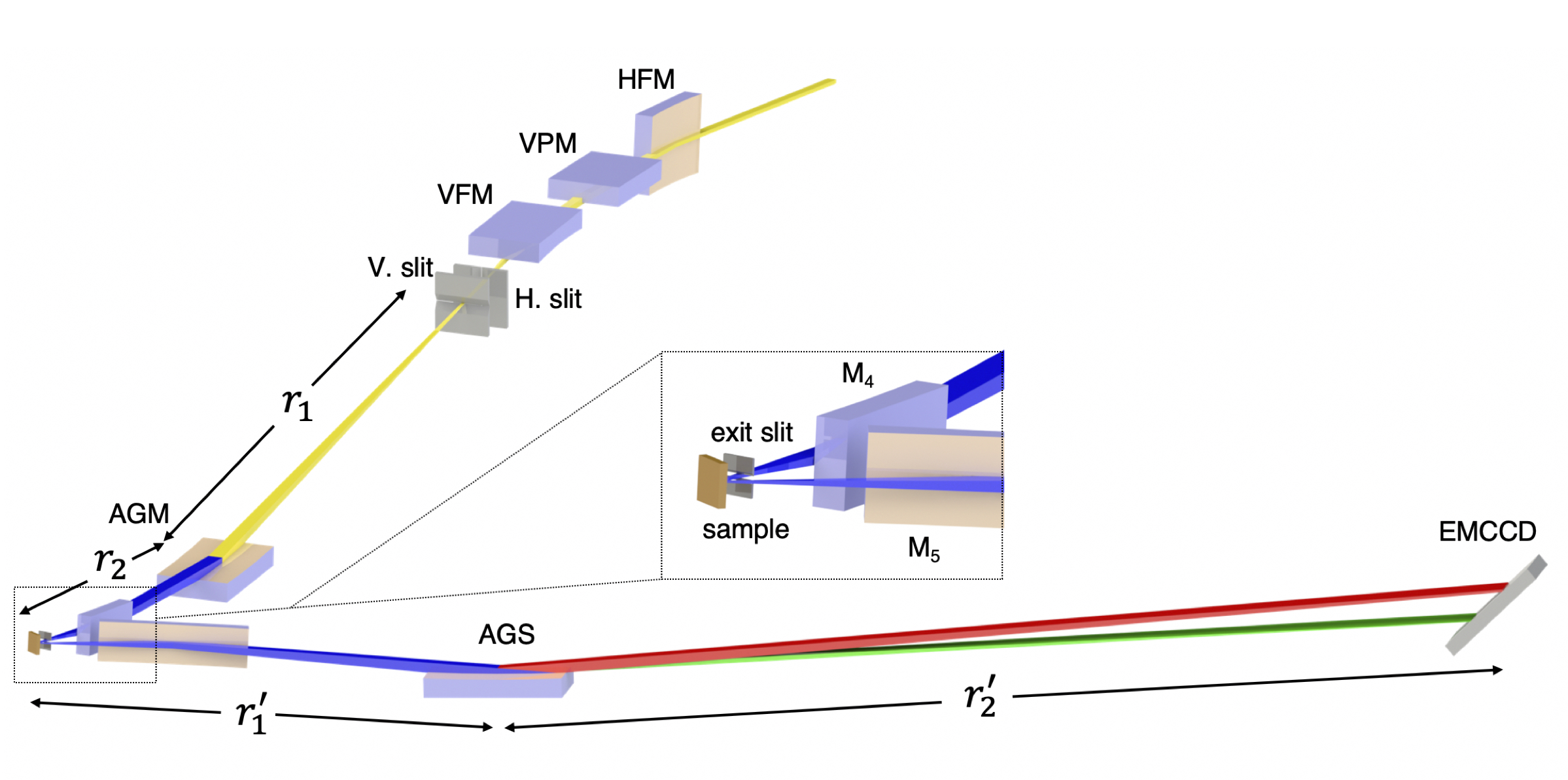}
\caption{Optical layout of AGM-AGS RIXS. Abbreviations of optical elements are defined in the text. Distances between optical elements are summarized in Tables 1 and 2.} \label{agmags}
\end{figure}

\begin{figure}
 \includegraphics[width=0.48 \textwidth]{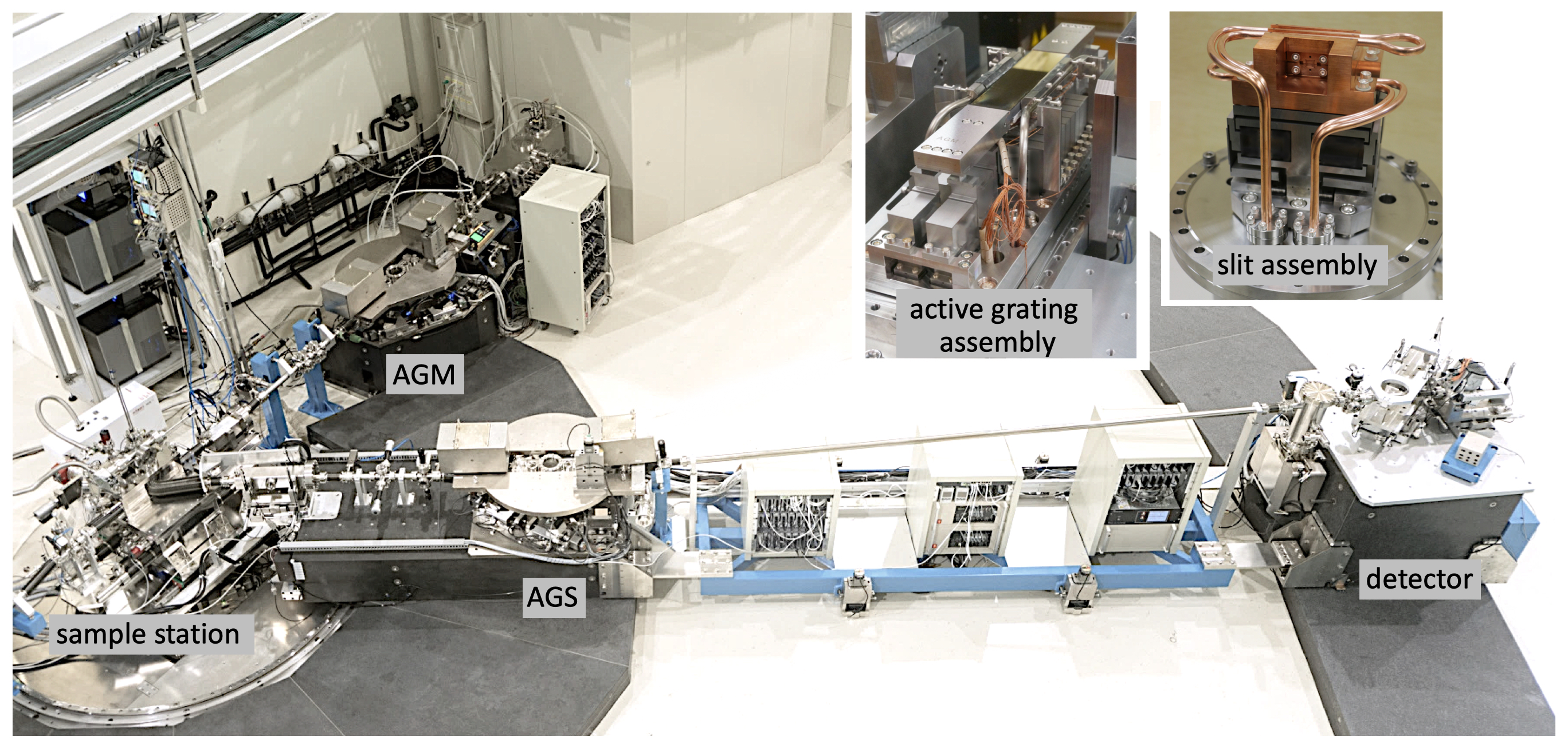}
\caption{Photographs of the RIXS setup installed at TPS beamline 41A. The AGS grating and the EMCCD detector are placed on two separate granite platforms. Through an air-cushion mechanism, the spectrometer can swing over a wide scattering angle from 17$^\circ$ to 163$^\circ$ about a vertical axis at the sample position.  Inset photographs show  the water-cooled active grating and slit assemblies.}\label{blpic}
\end{figure}

\subsection{Focusing optics and slits}

The RIXS branch includes the front end, focusing mirrors, slits, AGM grating, sample station, AGS grating and detector.  Figure \ref{agmags} illustrates its optical layout;  Figure \ref{blpic} shows the beamline photographs. The first optical element is a horizontal focusing mirror (HFM) located at 25.805~m from the center of the EPU tandem. After that, a vertical plane mirror (VPM) and a vertical focusing mirror (VFM) are located at 1.2~m and 2.4~m from the HFM, respectively.  The VFM focuses the photon beam onto a vertical slit, i.e., the entrance slit of the AGM, with demagnification 28.2.  The entrance-slit assembnly comprises two water-cooled diamond blades. Through a flexure-based high-precision positioning mechanism, the translation and the tilt of the two blades are controlled independently with four actuators with resolutions 0.002 $\mu$m and 0.02 $\mu$rad, respectively.  The opening of the entrance slit and its center can be varied continuously from 0 to 4000 $\mu$m.  The monochromator uses an active grating located 4 m after the vertical entrance slit to focus the incident soft X-rays vertically onto the sample located 2.5~m after the AGM grating.

The bandwidth of the soft X-rays dispersed from the AGM grating is selected with an exit slit made of tungsten carbide.  Through a translational stage driven with a UHV-compatible piezo motor, six discrete openings, i.e., 5, 10, 20, 50, 100 and 200 in units of $\mu$m, can be selected.  The exit slit placed 27.5~mm before the sample also sets the vertical beam size of the incident soft X-rays on the sample.  In the horizontal direction, soft X-rays are focused in two steps, first with the HFM which has demagnification 8.3, followed by a horizontal refocusing mirror mirror M$_4$ placed 0.6 m before the sample with demagnification 10.3. The designed horizontal beam size on the sample is 4.5~$\mu$m at FWHM.  The soft X-rays scattered from the sample are focused with another horizontal focusing mirror M$_5$ which has a collection angle 18~mrad to enhance the efficiency of the AGS.  The spectrometer uses an active grating located 2.5~m after the sample to focus the scattered soft X-rays vertically onto a two-dimensional (2D) detector located 5.5 m after the AGS grating.  The major parameters of the focusing mirrors are summarized in Table \ref{mirror}. 

\vspace{10mm}

\begin{table}
\caption{Major parameters of mirrors used at the RIXS branch of TPS beamline 41A. All mirrors have a Au-coated surface. HFM is with a Glidcop substrate; others are with a Si substrate. All values of $r_1$, $r_2$, and radius are given in units of m.} \label{mirror}

\begin{tabular}{llllll}
     
\\ \hline
  optic & type & $r_1$     &  $r_2$ & radius$^\ast$ & deviation angle ($^\circ$)  \\
\hline
HFM$^\#$   &   cylindrical      & 25.805   & 3.100      & 211      & 3 \\
VPM$^\#$   &   plane            & $\infty$ & $\infty$ & $\infty$ & 3  \\
VFM$^\#$   &   active   & 28.205   & 1.000        & 74       & 3  \\
M$_4$ &   plane-elliptical & 6.200      & 0.600      & 42     & 3  \\
M$_5$ &   plane-elliptical & 0.750     & 9.500       & 36       & 4 \\
\hline
\end{tabular}
\begin{tabular}{c}
$^\ast$ radius at the mirror center\\
\end{tabular}
\begin{tabular}{l}
$^\#$ HFM, VPM and VFM are also known as M$_1$, M$_2$ and M$_3$, \\respectively.
\end{tabular}

\vspace{3mm}
\end{table}

\subsection{AGM–AGS scheme}

The design of the monochromator and spectrometer is based on the energy-compensation principle of grating dispersion.  Incident X-rays from the entrance slit are diffracted, dispersed and focused onto the sample with the AGM grating; similarly, scattered X-rays are diffracted, dispersed and focused onto the 2D detector with the AGS grating.  The entrance and exit arms of AGM are,  respectively, $r_1$ and $r_2$; those of AGS are $r_{1}'$ and $r_{2}'$, respectively. The AGM-AGS scheme requires that the two gratings have an identical groove density $n_0$ at the grating center and that $r_{1}'$=$r_2$. As shown in Fig. \ref{agmags}, the inelastically scattered X-rays with the same energy loss but different incident energies have the same dispersion from the AGS, so implementing the energy-compensation principle of grating dispersion.  
The AGM-AGS scheme has two important features.  (1) The resolution and spectral-weight distribution of RIXS are insensitive to the incident bandwidth as long as the selected bandwidth is smaller than the core-hole lifetime width;  (2) The measurement efficiency is proportional to the selected bandwidth, because the energy-loss spectrum is the summation of the inelastic scattering excited by the incident photons within the bandwidth. Therefore, the AGM-AGS scheme can greatly enhance the efficiency of RIXS measurement while maintaining the energy resolution.

\subsection{Active gratings}

To implement the AGM-AGS scheme for high-resolution RIXS, two high-precision active gratings are required.  Using a multi-actuator bender, one can obtain a surface profile of grating with a high-degree polynomial to diminish defocus, coma and higher-order aberrations as well as to decrease the slope errors caused by thermal deformation and optical polishing.  In this RIXS beamline, each active grating is made of a varied-line-spacing plane grating mounted on a high-precision 25-actuator bender \cite{kao2019high}; its surface slope is monitored with an in-position long-trace profiler (LTP) \cite{lin2019development}.  Table \ref{tableagm} lists the optical parameters of the AGM and AGS gratings.

\vspace{10mm}

\begin{table}
\caption{Optical parameters of AGM and AGS gratings. Both gratings have a laminar groove profile with duty ratio 2:3 and groove depth between 8~nm and 6~nm with a position-dependent density $n(x) = n_0+n_{1}x$, where x is defined in Section 3.2.} \label{tableagm}

\begin{tabular}{lcc}
     
  \\\hline
  & AGM    & AGS   \\

\hline
Size (L$\times$W$\times$T) ($mm^3$)  &   186$\times$45$\times$10 & 186$\times$45$\times$10 \\
Substrate  &  Si & Si   \\
Coating  &  Au & Au  \\
Radius (m)  &  85-155 &  50-60 \\
Entrance arm (m)  &  $r_{1}$= 4.0 &  $r_{1}'$= 2.5 \\
Exit arm (m)  & $r_{2}$= 2.5 &  $r_{2}'$= 5.5 \\
$n_{0}$ (grooves $mm^{-1}$)  & 1200 &  1200 \\
$n_{1}$ (grooves $mm^{-2}$)  & 0.80 &  -0.32 (low-energy) \\
  &  &  -0.13 (high-energy) \\  
  \hline
\end{tabular}
\end{table}

\subsection{Optical tables and AGS rotation platform}

The implementation of high-resolution RIXS requires mechanical adjustments and supports with great precision and stability for all mirrors and gratings.  To fulfill these requirements, a high-precision, highly rigid and high-load optical table was designed and constructed for HFM, VPM, VFM, AGM and AGS. This optical table can support a UHV chamber for an optical element up to 1000~kg in weight with adjustments in all six degrees of freedom, i.e., three translations and three rotations. The resolution and repeatability of translation adjustments are 0.01~$\mu$m and 0.05~$\mu$m, respectively; those of rotation adjustments are 0.02~$\mu$rad and 0.1~$\mu$rad, respectively.
With comparable resolution and repeatability, an all-flexure-made high-precision optical table with adjustments in five degrees of freedom was designed and constructed for M$_4$, M$_5$ and the 2D detector.

To facilitate the variation of scattering angle, horizontal refocusing mirror M$_5$, the AGS grating and the 2D detector are mounted on a rotational platform composed of two movable air-cushioned granite blocks with a high-precision quickly detachable connecting bridge in between, as shown in Fig. \ref{blpic}. This design can minimize the ground micro-vibrational effects during data acquisition.  M$_5$ and the AGS grating are placed on one block, the 2D detector on the other.
During a rotation of the platform, the two blocks are connected with the bridge; the air gap between the granite block and the granite floor is kept at 30-$50~\mu$m.  Coupled with a unique UHV-compatible scattering chamber without a differential pumping, a wide scattering angle, 146$^\circ$, i.e., from 17$^\circ$ to 163$^\circ$, can be achieved.

\subsection{Sample manipulation}

\begin{figure}
\includegraphics[width=0.35\textwidth]{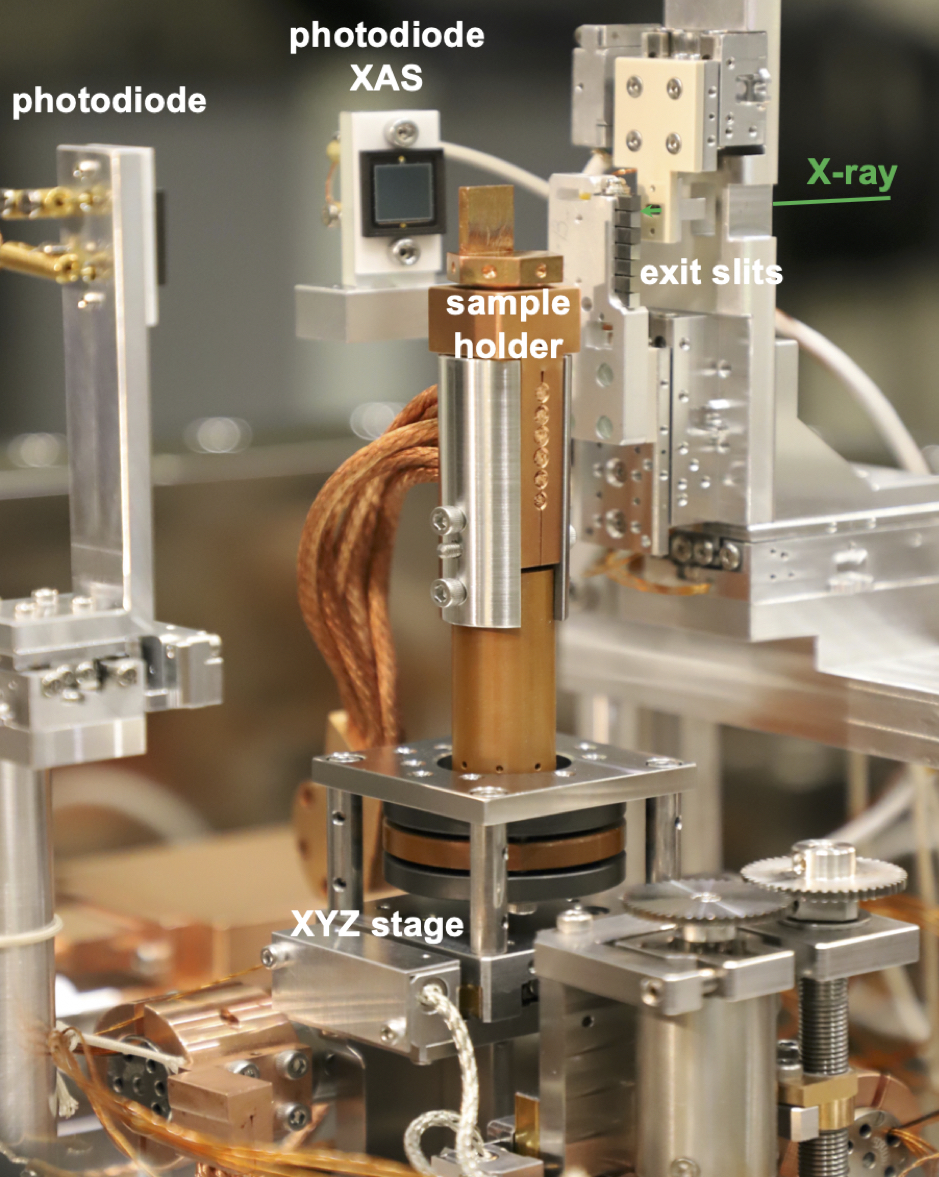}
\caption{Photograph of the sample manipulator inside the RIXS chamber. The sample xyz stage was developed using UHV compatible stepper motors and optical encoders. The sample position can be aligned with the incident X-ray beam with a resolution 1 $\mu$m. Two photodiodes are installed to measure the incident photon flux and fluorescence.}\label{xyz}
\end{figure}

The sample manipulator enables linear motions along three orthogonal directions with resolution 1 $\mu$m.  These motions are driven with a translational stage using UHV stepper motors and optical encoders.  Figure \ref{xyz} shows a photograph of this manipulator, which is mounted on top of an in-vacuum single-axis goniometer.  The sample holder made of oxygen-free high-conductivity copper is connected to a liquid-He cryostat through copper braids.  The sample holder is thermally isolated from the linear stage through a Vespel rod.  The sample can be transferred in vacuum through a load-lock chamber and can be cooled to 20~K.  In addition, the sample holder is electrically insulated from the cryostat, which enables us to perform measurements of X-ray absorption spectra (XAS) in the total-electron-yield mode.  To minimize the radiation-induced damage and saturation of the detector, two shutters have been installed, one before mirror M$_4$ and the other before the 2D detector. 

\begin{figure}
\includegraphics[width=0.45\textwidth]{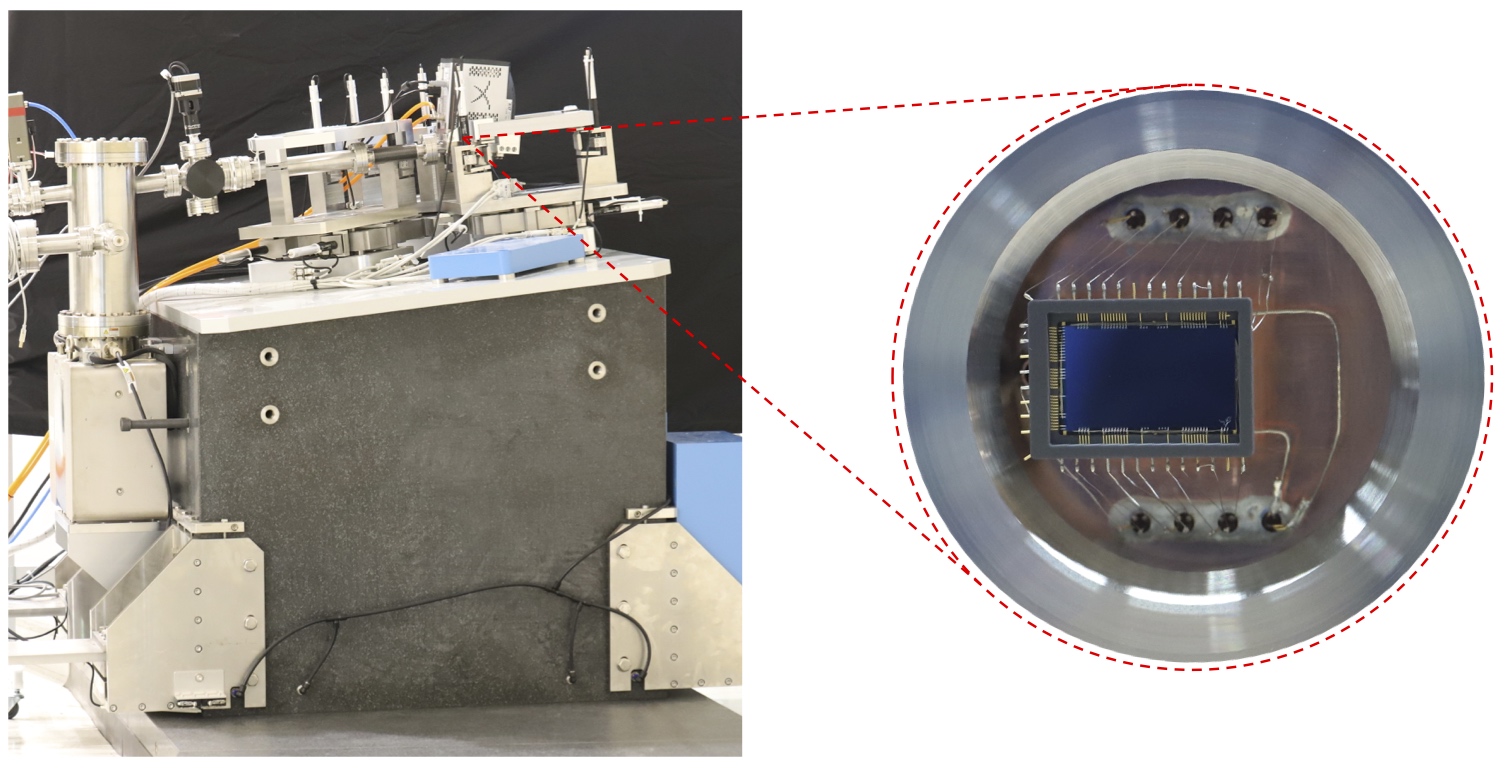}
\caption{RIXS 2D image detector. A photograph of the detector mounted on a granite floor is shown in the left. Details of a customized EMCCD sensor located inside the chamber marked with a red circle are revealed in the right.}\label{emccd}
\end{figure}

\subsection{Soft X-ray 2D detector}
The scattering cross section of RIXS is typically small and its signal is weak.  To achieve a detection scheme near photon counting, a detector of high sensitivity and low noise is required because only a few hundred electron-hole pairs can be generated with a silicon-based detector for soft X-rays.  Figure \ref{emccd} shows a photograph of a custom-made electron-multiplication CCD (EMCCD) detector without anti-reflection coating at grazing angle 10$^\circ$.  The pixel size of the EMCCD is 13.5~$\mu$m; the grazing incidence gives an effective pixel size 2.3~$\mu$m.  The EMCCD is back-illuminated to avoid absorption of photons by the integrated circuits on the front side of the wafer.  The detector head is cooled to $-85^\circ$~C.  Through the electron-multiplication mechanism, the signal-to-noise ratio is enhanced.  In addition, two four-jaw apertures are installed before both gratings to block stray light.

\subsection{Instrumentation control and user interface}

\begin{figure}
\includegraphics[width=0.48 \textwidth]{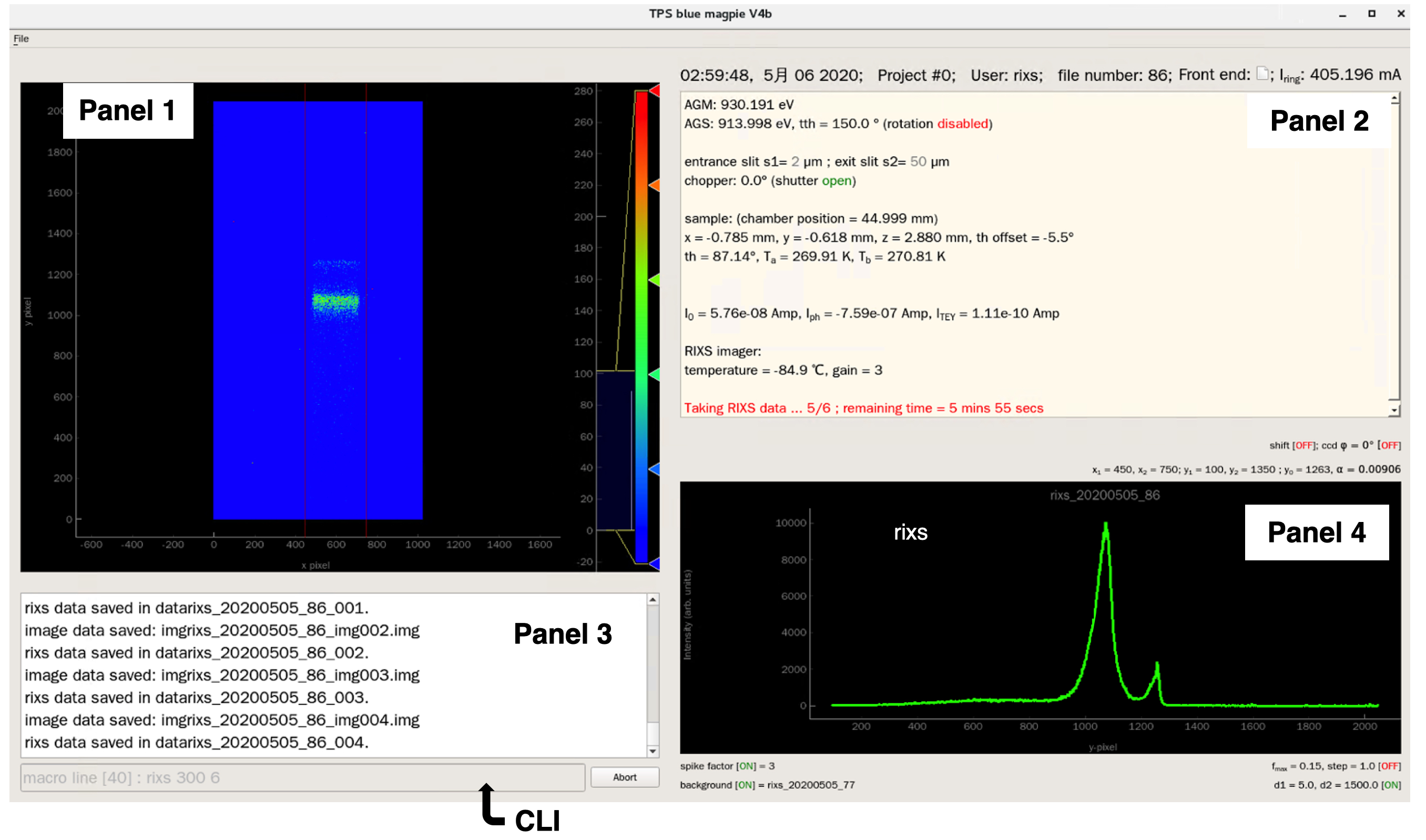}
\caption{Screenshot of the user interface, which is composed of four panels and one CLI.  Panel 1 shows the measured RIXS 2D image. Panel 2 displays the experimental parameters in real time.  Panels 3 and 4 show the command history and the 1D spectra for RIXS or XAS, respectively. }\label{magpie}
\end{figure}

The beamline instrumentation control for data accquistion is implemented on three levels: (1) the intrinsic commands of commercial devices, (2) the algorithm and software for a group control of commercial devices, (3) the user-level commands and interface.  Levels 2 and 3 are developed in house using FORTRAN and Python languages, respectively; the communications between them are based on EPICS tools.  Figure \ref{magpie} shows the screen of the user interface, which is composed of four panels and a user-oriented command-line input (CLI).  The first panel displays the RIXS 2D image recorded by the EMCCD.  The second panel displays the real-time status of measurement parameters, including the positions and orientations of optical elements, monochromator and spectrometer energies, sample rotation angle ($\theta$), scattering angle ($2\theta$), photon flux, sample temperature, shutter status etc.  The third panel shows the command history.  The fourth panel presents a graph of RIXS spectra, XAS, $\theta$-$2\theta$ angular scan etc.

\section{Commissioning results}

\subsection{Beam size}

\begin{figure}
\includegraphics[width=0.4 \textwidth]{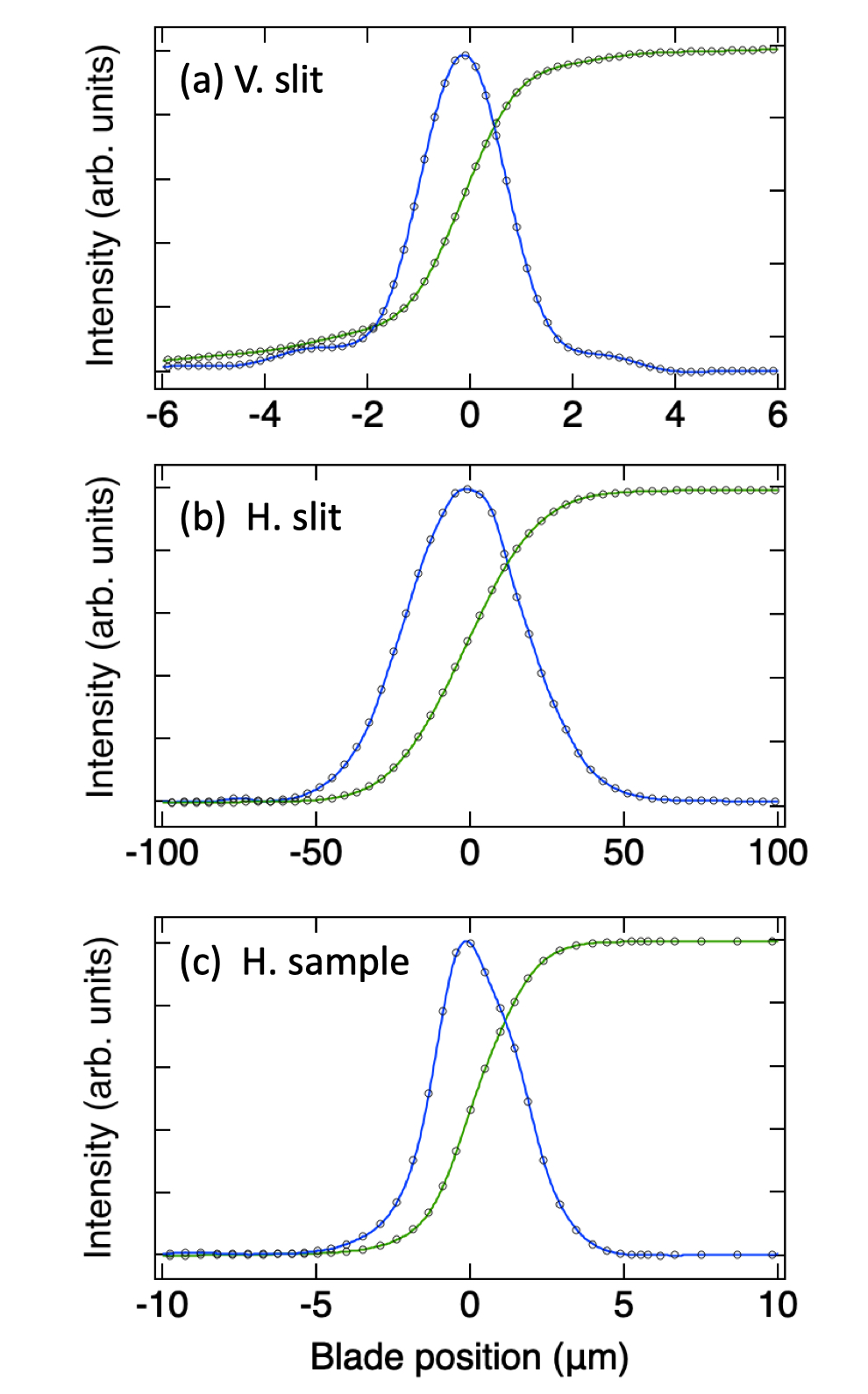}
\caption{Measurements of soft X-ray beam profile. (a) \& (b) Measurements of beam profile at the entrance slits for vertical and horizontal directions, respectively. (c) The horizontal beam profile of soft X-rays at the sample position.  The beam profile was obtained on detecting the beam intensity with a photodiode after a blade moving along the transverse direction.  The beam intensity from the photodiode as a function of the blade position is plotted with black circles along with a green curve which shows its Fourier filtered function; the beam profile after the differentiation of the photodiode signal is plotted with a blue curve for its Fourier filtered function.  The measured beam sizes of (a), (b) and (c) are 1.97, 44.38 and 3.23 $\mu$m at FWHM, respectively.}\label{bs}
\end{figure}

To achieve high-resolution RIXS measurements, a small soft X-ray source is required.  The knife-edge-scan method was used to measure the beam profile and to obtain the beam size at the slit positions.  Figures \ref{bs}(a) and \ref{bs}(b) show the measured beam-intensity profiles and their derivatives as a function of the knife-edge position of the vertical and horizontal entrance slits, respectively.  The derivatives of the measured profile reveal that the beam sizes at the vertical and horizontal entrance slits are 1.97 $\mu$m and 44.38 $\mu$m at FWHM, respectively, in satisfactory agreement with the designed values.   Figure \ref{bs}(c) plots the horizontal beam profile at the sample position; the focused beam size is 3.23 $\mu$m at FWHM in the horizontal direction.


\subsection{Adjustment of the grating surface profile}

\begin{figure}
\includegraphics[width=0.4 \textwidth]{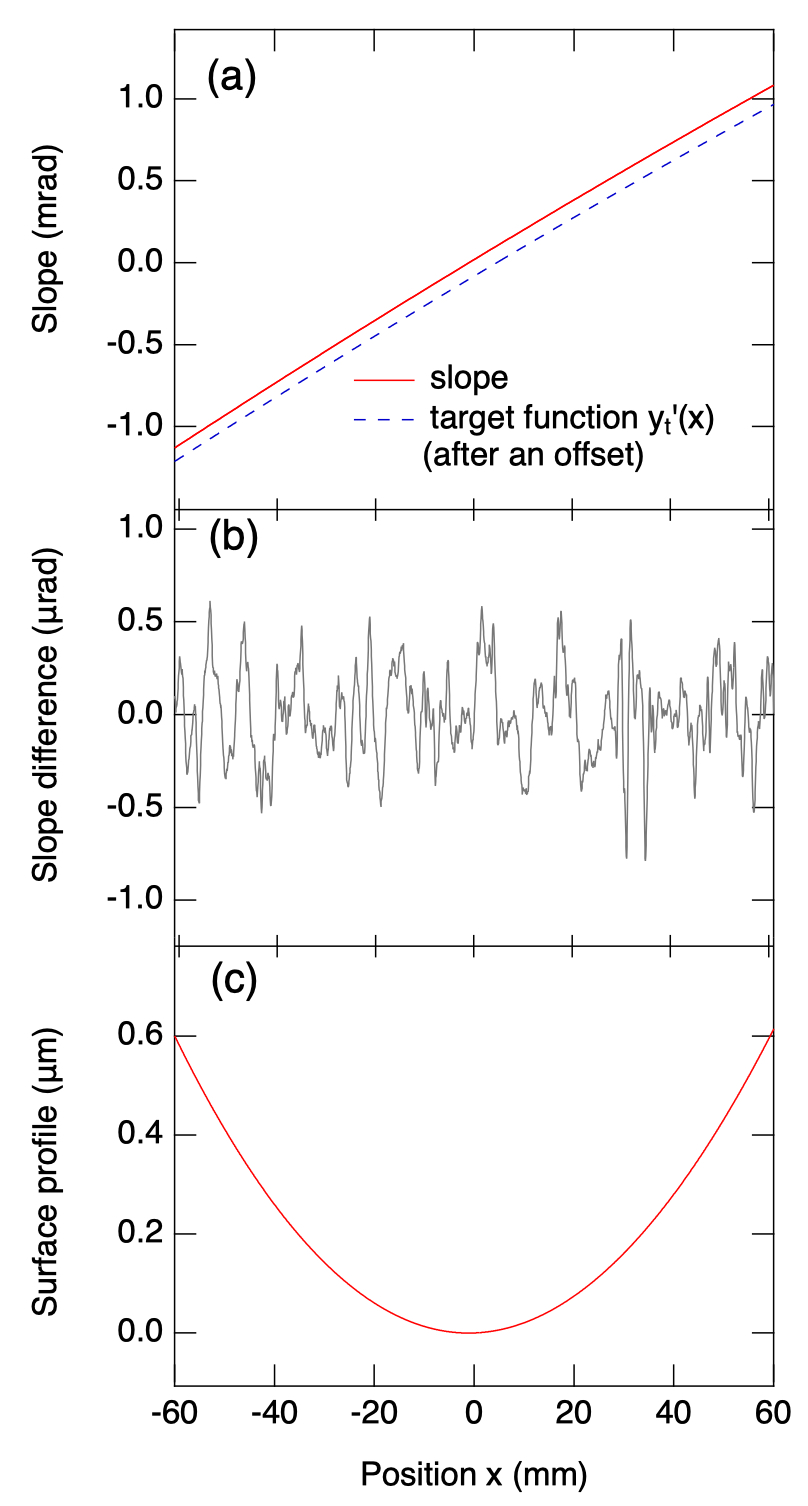}
\caption{Slope measurements of an active grating with an in-position  LTP. (a) Measured grating slope and its target polynomial slope function $y{_t}'(x)=c_{1}x+2c_{2}x+3c_{3}x^{2}+4c_{4}x^3$, in which $c_{1}=1.95{\times}10^{-5}$, $c_{2}=9.08{\times}10^{-6}$, $c_{3}=-3.85{\times}10^{-9}$ and $c_{4}=5.94{\times}10^{-14}$. The function $y{_t}'(x)$ is vertically offset for clarity.
(b) Difference between measured and target slope functions. The rms of this difference is $0.23~{\mu}$rad. (c) Profile function of the grating surface obtained from the integration of the measured slope, with the height at the grating center defined as zero.}\label{ltp}
\end{figure}

To operate the AGM-AGS optical system, the surface profile of each of the two active gratings must be adjusted to meet its target profile for a given incident photon energy.  The target profile y$_t$(x) is expressed as a polynomial function,
\begin{equation} 
 y_t (x) = \sum_{k=0} c_{k} x^{k},  \nonumber
\label{eqa1}
\end{equation}
in which $x$ is the position along the longitudinal direction of the grating and c$_k$ is the k$^{th}$ coefficient of the polynomial.  Our theoretical simulations indicate that a third-degree polynomial profile is sufficient to achieve an energy-resolving power 20,000, whereas a fourth-degree polynomial is necessary for 100,000.  To facilitate the adjustment of the profile of the grating surface, an in-position LTP system was used to measure the slope of this profile.  With the measured slope function, the target slope function and the set of actuator-response functions as inputs, one can use an iterative algorithm to deduce a set of actuator incremental values to adjust the surface profile to match the target profile.  The details of the adjustment are reported elsewhere \cite{kao2019high}.  Figure \ref{ltp} plots typical LTP measurement results, including the measured slope function and its target slope polynomial function, the difference between them and the measured surface profile of the grating.  The slope differnce between the measured and target functions was minimized to a root-mean-square (rms) value less than 0.25 $\mu$rad for both AGM and AGS gratings. Because of the limited precision and accuracy of the in-position LTP instrument, the obtained slope difference is larger than the slope error 0.1~$\mu$rad rms of the polished surface provided by the manufacturer.

\subsection{Resolution optimization}

\begin{figure}
\includegraphics[width=0.45 \textwidth]{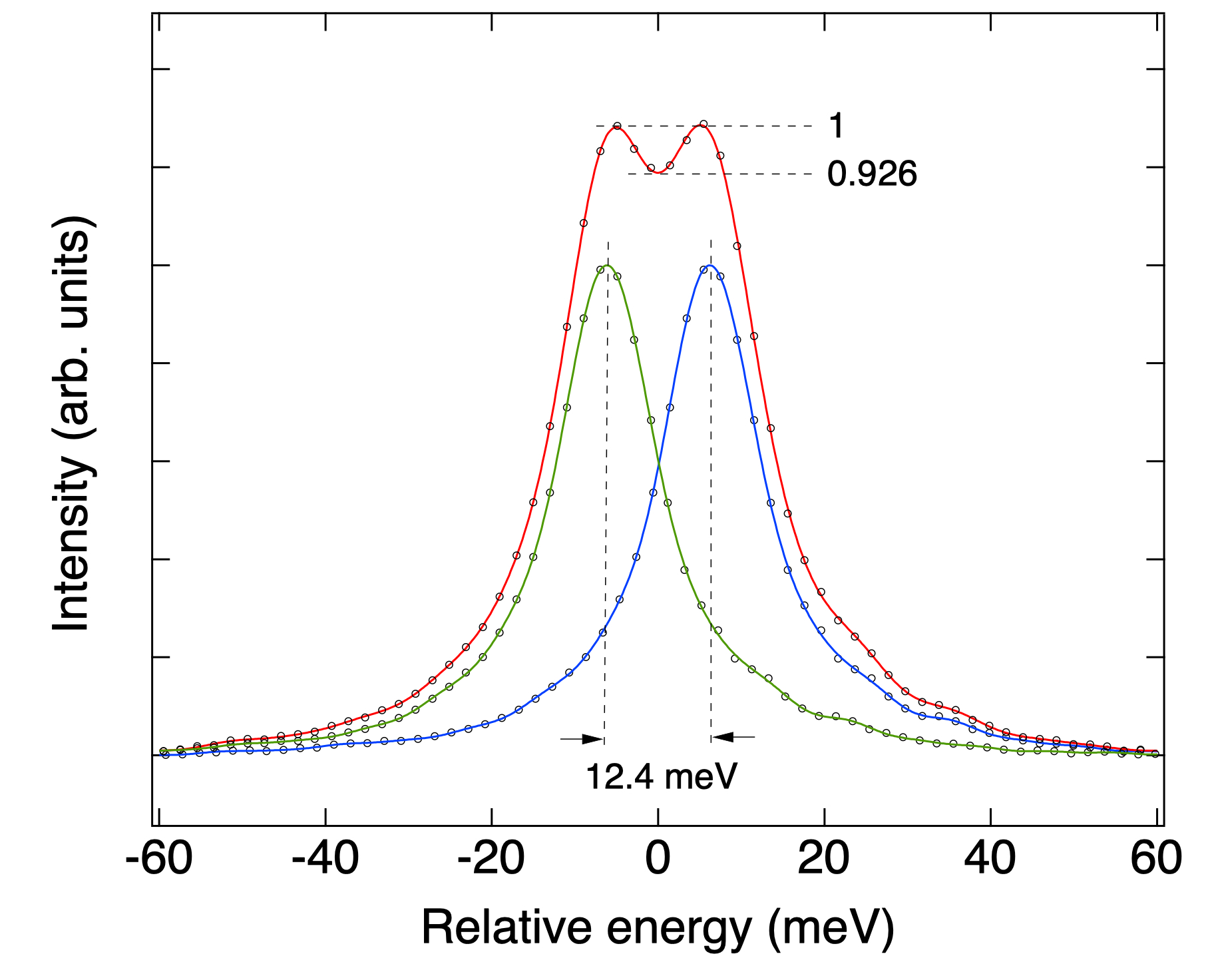}
\caption{Elastic scattering spectra of a W/B$_4$C multilayer measured in a near reflection condition at 530 eV. 
The measured scattering spectrum, which has a FWHM of 14.8~meV, is plotted with black circles along with a green curve which shows its Fourier filtered function.   The spectrum plotted along with a blue curve for its Fourier filtered function is a duplicate of this measured spectrum. When the separation between these two spectra is 12.4~meV, the core-to-wing ratio in their summation plotted together with a red curve for its Fourier filtered function is 0.926. The measured spectrum was obtained without applying any pixel elimination with an intensity threshold on the 2D image recorded in 1 sec.}\label{okedge}
\end{figure}

We measured the linewidth of elastic scattering from a W/B$_4$C multilayer (ML) to tune the energy resolution.  At the beginning, the grating surface profile was adjusted using the 25 actuators to match the initial target profile from the theoretical simulations of the designed beamline.  Because of the alignment imperfections of the beamline, various sets of polynomial coefficients, i.e., [c$_2$, c$_3$, c$_4$], were tested as the target profile to optimize the energy resolution.  
Figure \ref{okedge} shows the elastic scattering of 530 eV soft X-rays from the ML near a reflection condition with entrance slit 4 $\mu$m and exit slit 100 $\mu$m, which corresponds to bandwidth 0.5 eV of incident soft X-rays.  The best achieved spectral resolution was 12.4 meV, which is defined as the minimum separation between two spectral lines when they can be resolved through a criterion that the core-to-wing ratio is smaller than the ratio 0.9272 of two identical Gaussian functions separated by their FWHM.

\subsection{Photon flux on the sample}

As RIXS is a photon-demanding technique, a large photon flux on the sample is essential to achieve high-resolution measurements.  The AGM-AGS scheme allows us to increase the bandwidth of the incident photons on the sample while maintaining the energy resolution.  To verify this condition, we measured the energy resolution of the elastic scattering of 530 eV soft X-rays with exit-slit opening set at 50, 100 and 200 in units of $\mu$m.  The measured photon flux increased linearly from 3.6$\times$10$^{12}$ photons s$^{-1}$ to 1.3$\times$10$^{13}$ photons s$^{-1}$ with increased exit-slit opening.  The measured energy resolution remained nearly unchanged while the exit-slit opening increased.  These data are consistent with our previously results \cite{lai2014} and demonstrate that the AGM-AGS scheme is also applicable to high-resolution soft X-ray RIXS.

\subsection{Phonon excitations of superconducting cuprates}

The dressing of electrons by phonons in a material plays an important role in their novel electronic properties; in particular, the coupling between electrons and phonons in HTSC is pivotal yet remains under debate.  RIXS has the potential to measure the strength of the coupling of electrons to lattice excitations.  It has been shown theoretically that high-resolution RIXS provides direct, element-specific and momentum-resolved information about the electron-phonon coupling \cite{ament2011determining, devereaux2016directly}.  Here we present O $K$-edge RIXS measurements on La$_{1.88}$Sr$_{0.12}$CuO$_4$ (LSCO) to reveal its momentum-resolved phonon excitations.

\begin{figure}
 \includegraphics[width=0.45 \textwidth]{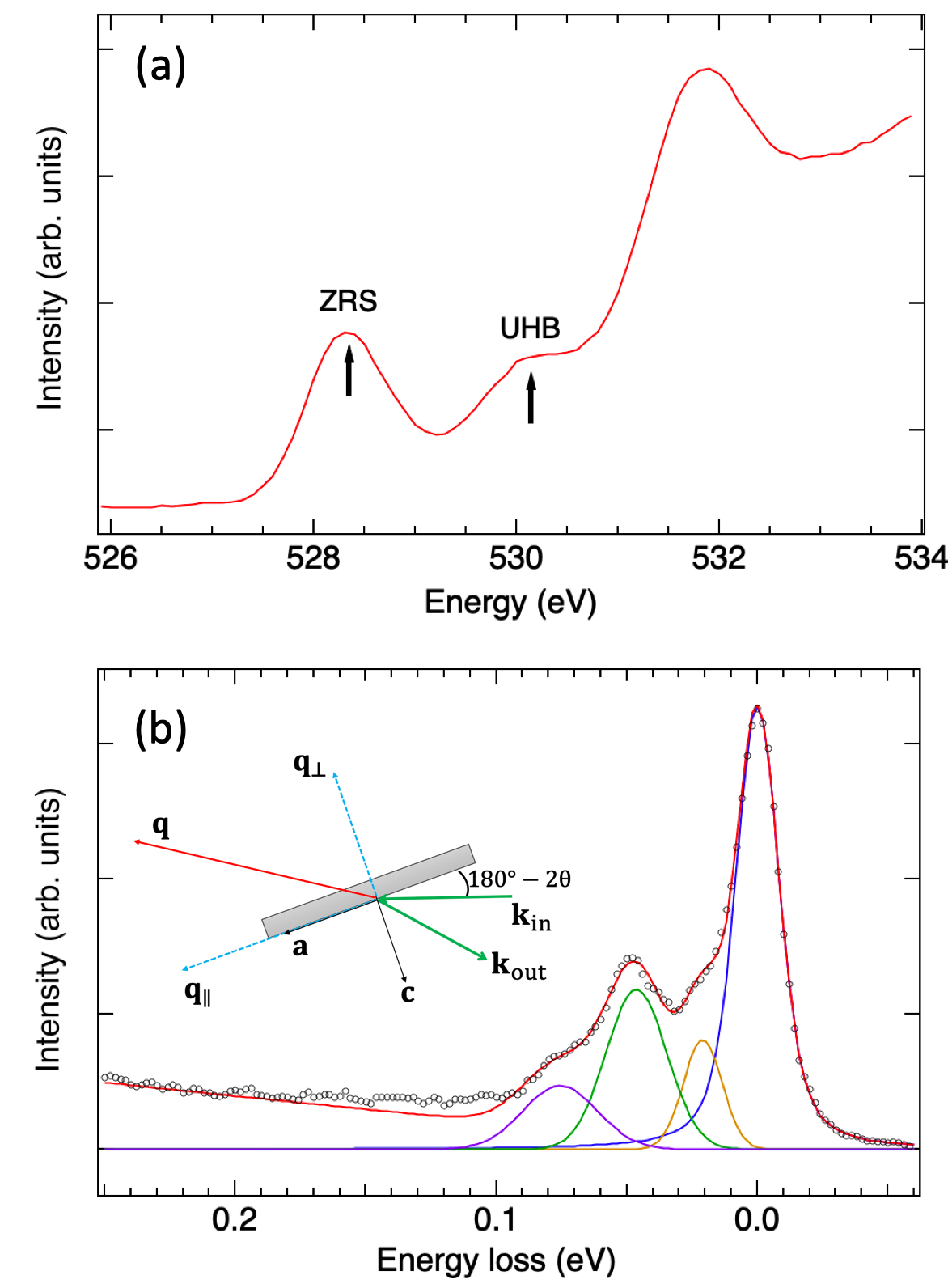}
\caption{(a) O $K$-edge XAS of LSCO at 23 K measured with the fluorescence yield mode with  X-rays of $\sigma$ polarization. (b) RIXS spectrum of LSCO with incident energy tuned to ZRS with $q{_\|}$~=~0.12~$\frac{2\pi}{a}$ and $q_\perp$~= 1.0~$\frac{2\pi}{c}$. The duration of exposure to record the RIXS data shown in the black circles was 2 hours. The red curve is the summation of the fitted components and a linear background function. Other color curves are fitted Voigt functions for elastic scattering and phonon excitations.  The inset illustrates the scattering geometry with {\bf{q}~=~\bf{k}$_{\rm in}-$\bf{k}$_{\rm out}$}, 
in which {\bf{k}$_{\rm in}$} \& {\bf{k}$_{\rm out}$} are incident and scattered wave vectors, respectively. {\bf a} and {\bf c} are crystallographic axes of LSCO. The scattering angle $2\theta$ was $150^{\circ}$.} \label{lsco}
\end{figure}

La$_{2-x}$Sr$_{x}$CuO$_4$ is a doped Mott insulator \cite{imada1998} composed of CuO$_2$ conducting layers with relatively weak coupling between the layers.  Its mother compound La$_{2}$CuO$_4$ is an insulating antiferromagnet, in which Cu ion has an electronic configuration $3d^9$ with a hole of symmetry ${x^{2} - y^{2}}$.  The system is a charge-transfer insulator because strong correlation effects split the conduction band into the upper and lower Hubbard bands.  These correlation effects manifest themselves in O $K$-edge X-ray absorption in which an electron is excited from the $1s$ core level to the $2p$ band.   The polarization-dependent XAS measurements revealed that the doped holes are distributed mainly throughout the O $2p$$_{x,y}$ orbital in the CuO$_2$ plaquette and are hybridized with the Cu $3d_{x^{2} - y^{2}}$ orbital to form a spin singlet $3d^{9}\underline{L}$ termed a Zhang-Rice singlet (ZRS), in which $\underline{L}$ denotes a ligand hole \cite{zhang1988effective, chen1992out}.  In the XAS of La$_{2}$CuO$_4$, there exists a prepeak at the absorption edge arising from the O $2p$ band hybridized with the upper Hubbard band (UHB) of Cu $3d$.  In a hole-doped cuprate, a lower-energy XAS feature resulting from ZRS emerges and grows linearly with hole concentration in the under-doped regime as plotted in Fig. \ref{lsco}(a) \cite{chen1991}.  That is, hole doping manifests itself in the spectral-weight transfer from UHB to ZRS as a consequence of electron correlations.  The existence of ZRS enables measurements of phonon excitations using O $K$-edge RIXS.  Figure \ref{lsco}(b) shows the RIXS spectrum of LSCO measured with incident X-rays set at 528.4 eV, i.e., the ZRS feature.  The inset of Fig. \ref{lsco}(b) illustrates the scattering geometry.  The in-plane momentum transfer $q{_\|}$ was along direction ($\pi$,~0) with {$q{_\|}$~=~0.12~$\frac{2\pi}{a}$}; the out-of-plane momentum transfer was $q_\perp$~= 1.0~$\frac{2\pi}{c}$, in which $a$ and $c$ denote the lattice parameters.  The RIXS spectrum in Fig. \ref{lsco}(b) shows an intense elastic peak and pronounced phonon excitations.  We fitted the measured RIXS spectrum with four peaks, i.e. an elastic-scattering peak and three phonon peaks, using Voigt functions.  The observed three phonon excitations originate from the half-breathing mode (75 meV), the A$_{1g}$ apical oxygen mode (45 meV) and the vibrational mode involved with La/Sr (20 meV).  These high-resolution RIXS measurements open a new opportunity to investigate the coupling between phonons and charge-density waves in HTSC.

\section{Summary and future plan}

We have designed, constructed and commissioned a high-resolution and highly efficient RIXS beamline based on the energy-compensation principle of grating dispersion.  The achieved resolving power was 42,000 at photon energy 530 eV with the bandwidth of the incident soft X-rays set at 0.5~eV.  With this high resolution, we observed three phonon excitations of LSCO.  
A new in-vacuum LTP instrument with a high precision of 0.005~$\mu$rad rms has been developed recently.  It will be installed in the beamline to greatly improve the accuracy of the measured slope function of the grating surface, aiming to reach ultra-high resolving power 100,000.
We also plan to install a polarimeter for the polarization analysis of scattered soft X-rays and to develop a high-spatial-resolution and highly efficient soft X-ray 2D detector.

\section{Acknowledgements}

We acknowledge the NSRRC staff for technical support; particularly we thank D. J. Wang for useful discussions. We also acknowledge L. Y. Chang for his help during the commissioning stage. Thanks are due to The Ministry of Science and Technology of Taiwan under grant No. 103-2112-M-213-008-MY3 and MOST 108-2923-M-213-001
 for providing partial support to this work.

\bibliography{41A.bib}
\end{document}